# Brownian Motion Captured with a Self-Mixing Laser

Kenju Otsuka, Koji Kamikariya, Takayuki Ohtomo, Seiichi Sudo* and Hironori Makino

*Department of Human and Information Science*
*Tokai University*
*1117 Kitakaname, Hiratsuka, Kanagawa 259-1292 Japan*

**Abstract**

We measured the overall motion of Brownian particles suspended in water by a self-mixing thin-slice solid-state laser with extreme optical sensitivity. From the demodulated signal of laser intensity fluctuations through self-mixing modulations by the interference between the lasing field and Doppler-shifted scattered fields from Brownian particles, it was found that the changes over time in random walks of small particles suspended in water result in different overall dynamics when the field of vision for particles seen by the laser beam (scale of the observation) is changed. At a small focal volume of the laser beam, in which the relevant diffusion broadening is observed, the overall motion which can be represented by the motion of a "virtual" single particle whose velocity obeys a double-peaked non-Gaussian distribution function. The fast motion of a virtual particle, featuring fractional Brownian motion, which leads to a pure random walk with increasing the observation time, has been characterized in terms of mean-square displacements of a virtual particle.

## 1. Introduction

In disciplines ranging from fluids through ecosystems, chemistry, and electronics to finance systems, we find phenomena associated with non-Gaussian intensity probability distribution functions that have 'long tails' or reflect 'self-similar' (i.e., non-stationary) processes. This family of distributions was discovered by Levy [1] and shown to be applicable to very many phenomena in a broader new interdisciplinary field by Mandelbrot [2]. This paper deals with a novel approach to the measurement and presentation of one such phenomenon discovered in small particles in Brownian motions, that is the prototypical example which exhibits non-stationary behavior, i.e., random walk. Our motivation behind this study is to capture an intriguing dynamic behavior of Brownian particles, which appears as an overall motion of independent random walks depending on the scale of observation. Such a study would provide a conceptually new insight into the general Brownian motions.

The coherent nature of laser light has been applied in dynamic light-scattering (DLS) method [3] for characterizing motions of small particles in suspension, including gasses, liquids, solids, and biological tissue [4-6]. DLS approach toward measuring diffusion broadening of scattered light from moving small particles, is helpful to extract useful information on particles in Brownian motions and particle sizing. In conventional DLS, there are two methods: In the first method, an intensity fluctuation of scattered light passing through a small pin-hole, which represents beat signals of Doppler-shifted fields scattered by different particles, is measured. Then, an autocorrelation function is calculated from long-term experimental time series. From the calculated autocorrelation function, the distribution of particle sizes can be analyzed. In the second method, beat signals between a local oscillator light field and a scattered light field are measured using an optical interferometer, in which a frequency shifter is introduced in one arm to create the frequency-shifted field while the un-shifted beam in another arm acts as a local oscillator field. In this heterodyne detection scheme, the spectral broadening is measured by a spectrum analyzer [7].

On the other hand, we developed so far the real-time nanometer vibration measurement system with extreme optical sensitivity using microchip solid-state lasers. In this scheme, the laser itself is intensity-modulated by beat signals between a lasing field (i.e., un-shifted local oscillator) and a frequency-shifted scattered field through the interference of two fields [8,9]. So, we do not need an optical interferometer and furthermore high optical sensitivity is ensured due to the enhanced self-mixing modulation effect which is proportional to large fluorescence-to-photon lifetime ratios in microchip lasers. Therefore, this self-aligned self-mixing photon correlation spectroscopy using backscattered light is expected to make easy particle sizing possible and give us a way to capture the overall motion of particles in real-time, each of which independently scatters the laser light. The distinct advantage of our self-mixing scheme is that we can demodulate the laser intensity fluctuations through a simple frequency-modulated (FM)-wave demodulation circuit [8,9].

In this paper, we demonstrate that quick particle sizing and determining distribution of size of particles in suspension are possible from analyses of power spectra of modulated signals exhibiting diffusion broadening obeying the Einstein-Stokes relation. From demodulated signals, it is shown that when the field of vision of particles by the laser beam is adjusted such that the relevant diffusion broadening is observed, random sequences featuring self-similarities appear in the overall movement of individual particles in the short time scale of observations. We observed non-Gaussian, but stationary velocity probability distributions for the overall movement.

## 2. Self-mixing laser measurement of Brownian particles

### 2.1 Experimental setup of self-mixing laser spectroscopy of Brownian particles

The experimental setup is shown in Fig. 1, which is similar to that used in refs. [8,9]. The laser was a laser-diode (LD) pumped 0.3-mm-thick $LiNdP_4O_{12}$ (LNP) laser with a coating of mirror on each end and operating at the wavelength of $\lambda$ = 1048 nm. The collimated beam from the LD (wavelength: $\lambda_p$ = 808 nm) was passed through a pair of anamorphic prisms and then focused on the LNP crystal by a microscope objective lens. The greater portion (96 %) of the output light was passed through an iris, frequency-shifted by two acousto-optic modulators (AOMs), and then delivered to the scattering cell. The fused quartz cell was filled with water that contained spherical polystyrene latex particles (STADEX: at 1% density).

Microscope objective lenses with different numerical apertures (NA = 0.25, and 0.4) were used to focus the laser beam such that the focal plane (beam waist) was within the cell, where a depth from the cell's inner surface (wall), $d$, was varied. The quartz cell's surface was slightly tiled to suppress the light feedback from the surface. Changing the modulation frequencies of the up-shift and down-shift AOMs produced a shift in optical carrier frequency of $2f_s$ = 2 MHz at the end of the roundtrip. The rest (4 %; 80 µW) of the output light was detected by an InGaAs photoreceiver (New Focus1811: DC-125 MHz), and the electrical signal produced by this device was fed to further electronic devices; rf spectrum analyzer (Tektronix 3026: DC-3GHz), digital oscilloscope (Tektronix TDS 540D: DC-500 MHz), and FM receiver. The laser's threshold level of pump power was 30 mW and its slope efficiency was 40 %.

### 2.2 Lorentz broadening and particle sizing

The self-mixing effect is produced by interference between the laser field and the Doppler-frequency-shifted field fed back from the moving particles to the laser; this led to intensity modulation of the laser at the beat frequency between the two fields [8,9]. The short-cavity configuration of our thin-slice solid-state laser led to a photon lifetime six orders of magnitude shorter than the 120-µs fluorescence lifetime. This compensates for the extremely weak optical feedback condition and gives us a means for the real-time examination of dynamic light scattering by particles in Brownian motion; that is, detection of the frequency-modulation-driven variations in the intensity of the output laser light at beat frequencies between a lasing (i.e., un-shifted local oscillator) and scattered fields from particles, $2f_s + f_D(t)$ ($2f_s$: carrier frequency).

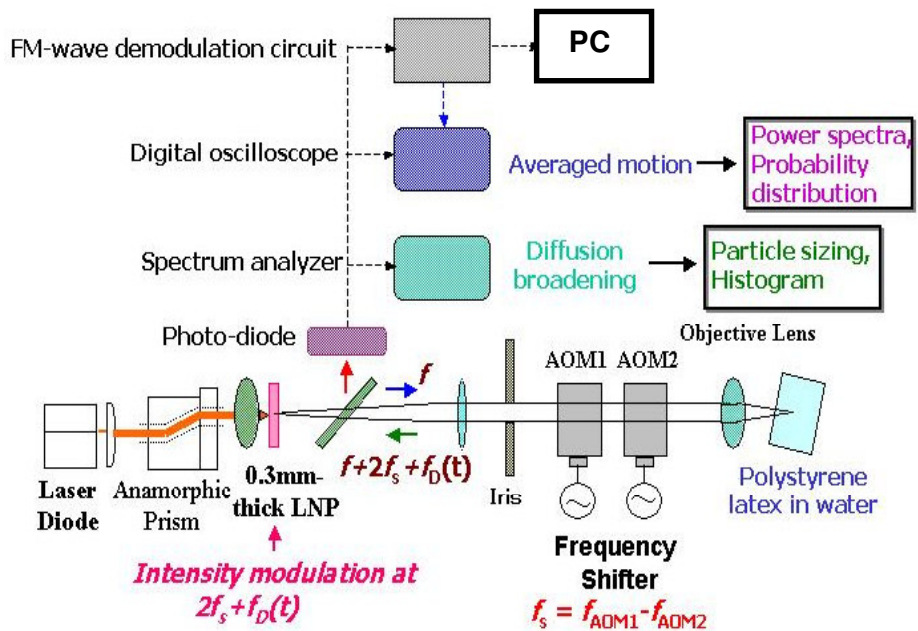

Fig. 1. Experimental setup.

Figure 2(a) shows an example of modulated output waveforms, with a magnified view of the waveform shown as the inset, where 207-nm-diameter particles were used. Note that a large fluctuation in the envelope of the 2-MHz carrier wave amplitude is a result of the fluctuation over time in the number of particles coming in and out of the region on which the laser beam is focused, since the modulation depth (i.e., feedback ratio) depends on the number of scattering particles. Let us examine power spectra of modulated signals. Figure 2(b) shows power spectra obtained for 107-nm, 207-nm, and 458-nm particles, where a microscope objective lens of NA = 0.25 was used. Each power spectrum was obtained by averaging 100 traces from the spectrum analyzer.

We could perform an excellent particle sizing for diluted sample with a $10^{-3}$ wt.% particle density with the present self-mixing laser.

### 2.3 Distribution of particle sizes

We have performed successful evaluations of *average* diameters of polystyrene latex spheres of different diameters (107 nm, 207 nm and 458 nm) within 5% accuracy by averaging of many (e.g., 100) power spectra of modulated signals like Fig. 2(b). Here, let us discuss about measurement of particle distributions by using the present self-mixing scheme.

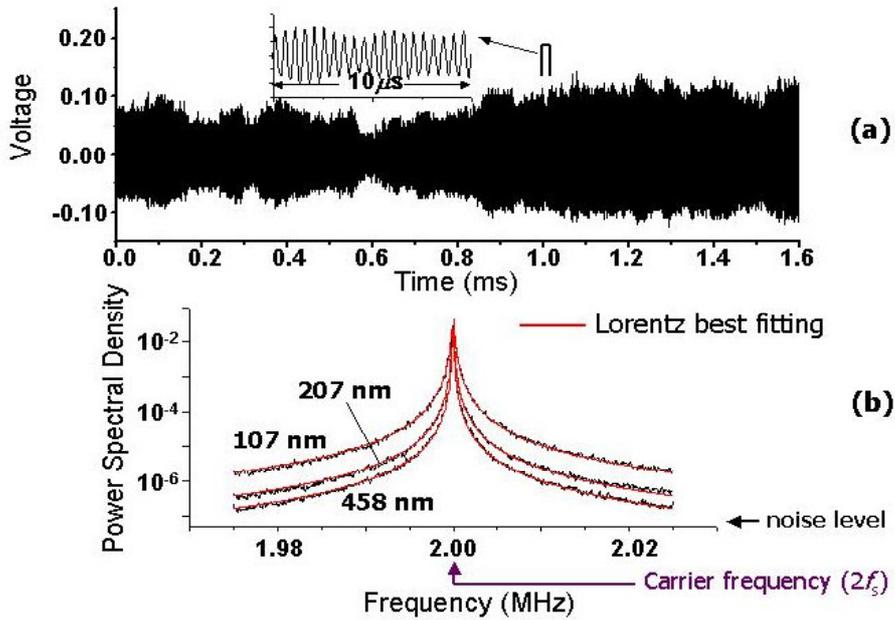

Fig. 2. (a) Modulated output waveform, (b) power spectra.

It is well known that the frequency spectrum of the light scattered by a Brownian particle show the Lorentz profile [3] according to the fluctuation-dissipation theorem. When each particle is moving independently and the scattered field amplitude from each particle is constant, the power spectral densities of modulated signals shown in Fig. 2(b) are expected to be fitted by the following Lorentz curve similar to that in the case of heterodyne detection using interferometer [7]:

$$I(\mathbf{k},\omega) = A + B(\mathbf{k}^2 D)/[(\omega - 4\pi f_s)^2 + (\mathbf{k}^2 D)^2], \quad D = k_B T/3\pi\eta a \quad (1)$$

Here, $D$ is the diffusion constant, $a$ is the diameter of the Brownian particles, $\eta$ is the liquid medium's coefficient of viscosity, $k_B T$ is Boltzmann's factor, and $\mathbf{k}$ is the wave vector. The observed rf power spectrum provides a good representation of the frequency spectrum of laser light scattered by independent particles in Brownian motion. The half-width at half-maximum frequency width, $(4\pi n/\lambda)^2 D \sin^2(\theta/2)$ ($\theta$: scattering angle; $\theta = \pi$ in our case), as estimated from the Lorentzian curve of best fit obtained by the fitting software shown in Fig. 2(b), yielded the particle diameter of 103 nm, 203 nm and 420 nm for different particles, respectively.

From FFT (Fast Fourier Transformation) of averaged one-side power spectra of modulated signals, we calculated the "net" normalized second-order autocorrelation function $G^{(2)}(\tau)$ which is related to the first-order autocorrelation function $g^{(1)}(\tau)$ through Siegert relation as [10, 11]:

$$g^{(1)}(\tau) = \int G(\Gamma) \exp(-\Gamma\tau) d\Gamma = [G^{(2)}(\tau)]^{1/2} \quad (2)$$

$$G^{(2)}(\tau) = [g^{(2)}(\tau) - C]/\beta, \quad (3)$$

with $G(\Gamma)$ being the normalized distribution of linewidth. C is he background which can be measured at large delay times $\tau$. We assume $\beta$ to be an unknown parameter in the data fitting procedure. The "net" signal autocorrelation function $G^{(2)}(\Gamma)$ calculated from the averaged power spectrum for 207-nm particles is shown in Fig. 3(a). By using the histogram method [11], we calculated the distributions of linewidth $G(\Gamma)$ and particle size from $G^{(2)}(\Gamma)$. Results are shown in Figs. 3(b) and 3(c), respectively, where the averaged particle size was 204 nm. Similar results were obtained for particles of 107-nm and 458-nm diameters.

We are now developing a statistical analysis tool for calculating a histogram of particle sizes directly from a measured power spectrum especially for mixed particles with different sizes.

particle along the laser axis, $v_x(t)$, whose instantaneous velocity is given by the sum of instantaneous velocities along the laser axis of individual particles, each of which has its own velocity vector over times. A super-heterodyne method with a central frequency of 10.7 MHz was employed in FMD in Fig. 1 to demodulate the FM wave; the amplifier had a gain of 20 dB and a 3-dB bandwidth of 111 kHz. An example demodulated output voltage is shown in Fig. 4(a) [12].

Note that the autocorrelation function and the corresponding

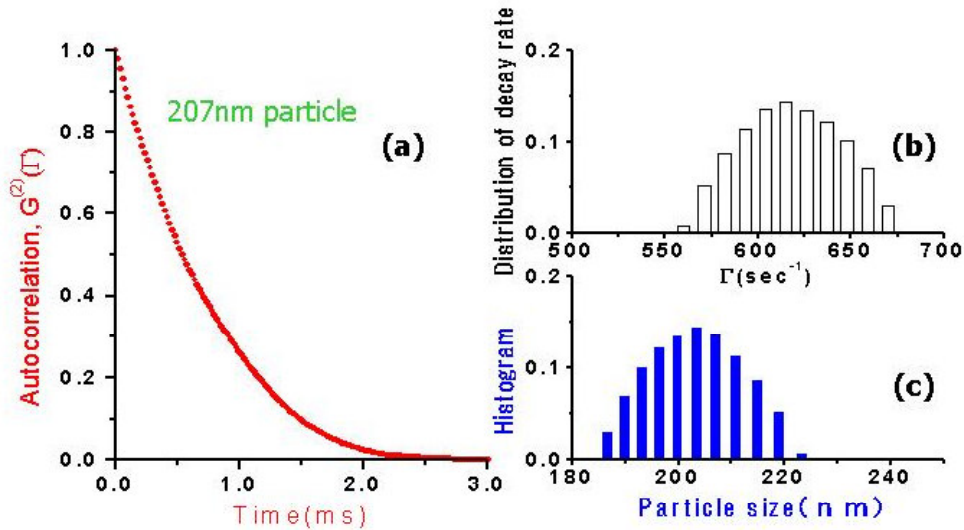

Fig. 3. (a) Normalized "net" autocorrelation function $G^{(2)}(\Gamma)$ calculated from the averaged power spectrum of modulated signal for 207-nm polystyrene latex in water assuming $\beta = 1$. (b) Probability distribution of linewidths $G(\Gamma)$. (c) Histogram of particle sizes.

## 3. Statistical properties of demodulated signals

### 3.1 Random sequence of Lorentz-type spectrum

Next, let us investigate an overall motion of Brownian particles within the scale of observation by the laser beam by demodulating the laser output intensity. A demodulated signal voltage, $V_o(t)$, is considered to be proportional to the velocity of a single "virtual" power spectral density exhibit an exponential decay and random sequences possessing Lorentz type of spectrum, as shown in Figs. 4(b)-(c), respectively where the curves of best fit are indicated.

The roll-off point from the $f^0$-type toward the asymptotic $f^{-2}$ behavior is found to shift to the low-frequency side with increasing the diameter of particles in suspension as shown in Fig. 5.

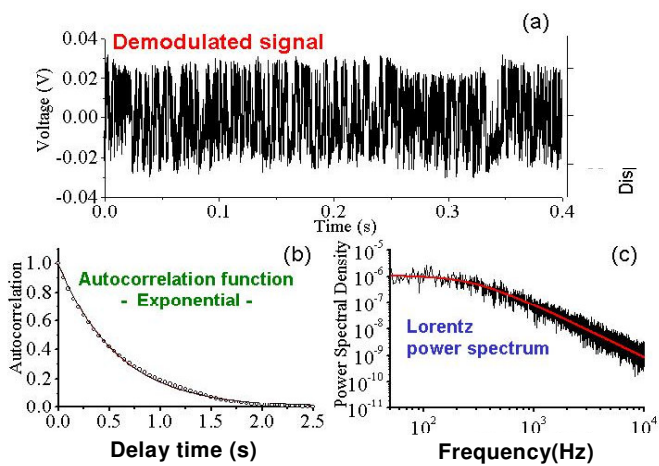

Fig.4. Demodulated signals. (a) Intensity wave form for 207-nm particles. (b) Autocorrelation function. (c) Power spectra.

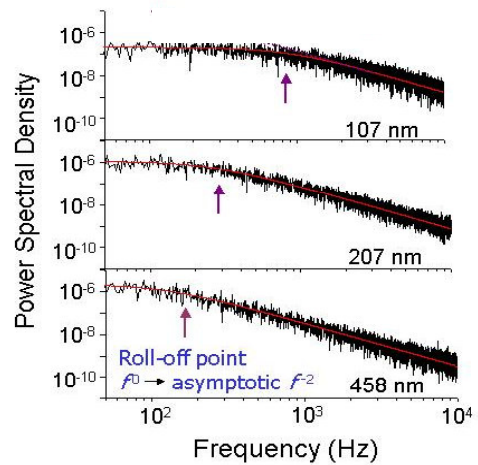

Fig. 5. Power spectra of demodulated signals for particles with different diameters.

Here, the focal volume of the laser beam, $V_s = 2\pi w_0 f_D$ ($w_0$: focused beam spot size, $f_D = \lambda/2(NA)^2$: focal depth), contained sufficient number of particles, i.e., N ~ 100, 1000, 8000 for 458-, 207- and 107-nm particles, respectively.

The Lorentz-type of statistical nature of the demodulated signals, i.e. velocity fluctuations, which is considered to be derived from the Langevin equation, implies that a virtual particle behaves just like a single Brownian particle.

### 3.2 Probability Distribution of Virtual Particle Velocity

The Lorentz-type of statistical nature of the demodulated signals, i.e. velocity fluctuations, shown in Fig. 5 implies that a virtual particle behaves just like a single Brownian particle which obeys the Langevin equation, in which the roll-off frequency, i.e., bandwidth, decreases with increasing the diameter. However, the velocity probability distribution function was found to strongly differ from Gaussian distribution functions as expected in particles in Brownian motions. Example results for particles with different diameters are shown in Fig. 6. The output voltage (i.e., velocity) fluctuation indicates non-Gaussian probability distributions featuring double peaks. While, the velocity ($v_x$) probability distribution of Brownian particles has been re-examined to be Gaussian [13].

In our self-mixing DLS scheme with a focused laser beam < 100 μm diameter, the field of vision of particles seen by the laser beam is strongly limited. Therefore, particles moving across the laser axis with fast speeds will disappear from the field of vision limited by the focused beam diameter. This implies that the probability around $v_x(t) = 0$ tend to be decreased as shown in Fig. 6. In our experiment with microscope objectives, "tails" in the velocity

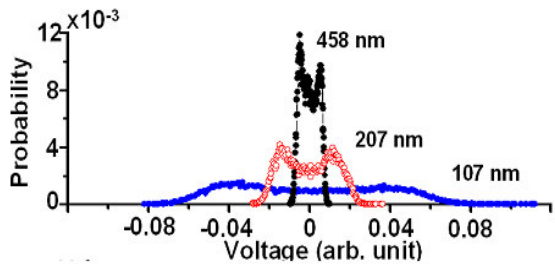

Fig. 6. Velocity probability distribution function of a virtual particle.

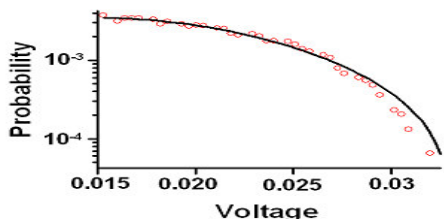

Fig.7. Truncation of a tail of the Gaussian distribution.
207-nm particles in Fig. 6. Solid curve: Gaussian function

Gaussian distribution functions are also truncated as seen in Fig.7, because particles moving along the laser axis with fast speeds will also disappear from the field of vision limited by the focal depth.

With increasing a beam-waist diameter and a focal depth much larger than those attained with a microscope objective lens by using a conventional lens, double-peaked distribution function was not observed in the demodulated output and it was replaced by the Gaussian function as expected, in which the modulated output signal was weakened due to the decreased numerical aperture.

## 4. Mean-squared displacement of a virtual particle

Let us examine mean-squared displacements of a virtual particle as a function of the time scale of observations. Figure 8(a) show motions of a virtual particle, $x(t) = \int v_x(t)\,dt = C \int V_o(t)\,dt$ (C: proportionality constant) which was directly obtained through a LabView software in the personal computer (PC), and the corresponding power spectra are shown in Fig. 8(b). We used the Hilbert transformation of the modulated output waveform and phase-sensitive detection in the PC to determine C and calculate the real displacement, $x(t) = \lambda\,\Delta\Phi(t)$, where $\Delta\Phi$ is the phase difference between the 2-MHz carrier (reference) wave and the modulated wave, as calculated from Gabor's analytic signal [8,9].

It is mathematically obvious that the power spectra of the corresponding motion of a virtual particle are given by those of velocity variations divided by $\omega^2$ ($\omega$: angular frequency), in which the roll-off frequency from the $f^{-2}$-type toward the asymptotic $f^{-4}$ behavior shifts to the lower frequency side as the diameter increases corresponding to Fig. 2(b). Fitting curves are shown in Fig. 8(b).

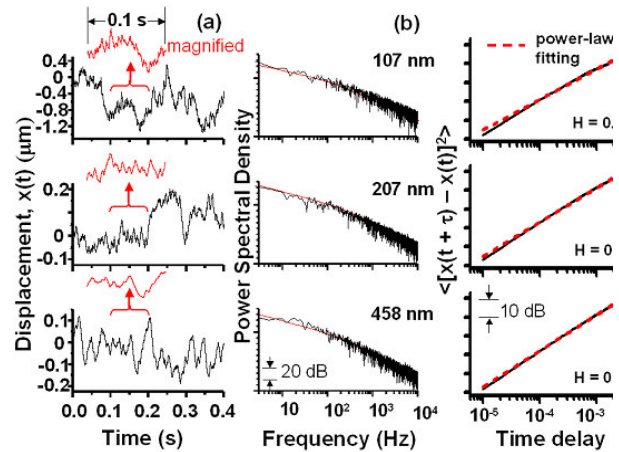

Fig.8. Temporal displacement of a virtual particle, power spectrum and mean-squared displacement as a function of observation (i.e., delay) time.

Figure 8(c) shows the corresponding structure function, $S(\tau) = \langle[x(t + \tau) - x(t)]^2\rangle$, for different diameter particles, where $\tau$ is a variable time delay (i.e., observation time). The ensemble average of 200 data extracted from the time series was calculated for each $\tau$. Many timescales of fluctuations are present corresponding to the power spectra shown in Fig. 8(b) because of the relation $f^{-\alpha}$; $\alpha = 2H + 1$, where H is the Hurst exponent. (0 < H < 1). Note that the power law, $S(\tau) \sim \tau^{2H}$, nearly holds over the relatively short time scale of observations for all cases, as depicted by dashed lines. The Hurst exponent, which represents the degree of persistence or long-range dependence, increases as the diameter increases, as depicted in the figures. This result reflects the stronger suppression of higher frequency components for larger particles and the dynamics tends to exhibit stronger self-similar properties, i.e., fractional Brownian motions with H > 0.5 [2], in which the increments of the process are positively correlated. As for 458-nm particles, in particular, H = 0.955, implying strong self-similarities. Note that the velocity probability distribution function was found to exhibit the non-Gaussian nature in such short observation time scales

On the other hand, the mean-squared displacement of Brownian particles is given by $S(\tau) = 2D\tau - 2D(m/\gamma)[1 - \exp\{-(\gamma/m)\tau\}]$, from the Langevin equation, where $\gamma = 6\pi a \eta$ is the dissipation coefficient.

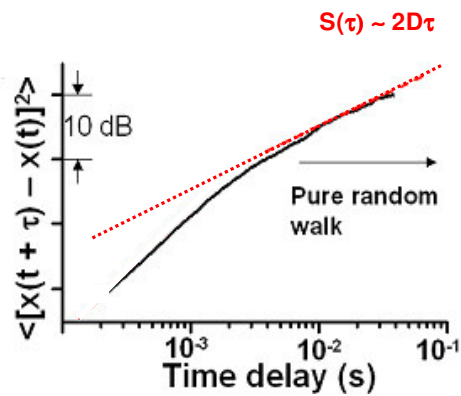

Fig. 9. Mean-squared displacement of a virtual particle for an increased observation time. 1-wt.% 108-nm particles in water.

Assume relevant $\gamma$ and m values for real Brownian particles with 107-, 207-, and 458-m diameters, the exponential decay term vanishes and pure random walks, i.e., $S(\tau) = 2D\tau$ should appear in the observation time scale shown in Fig. 8, $\tau > m/\gamma = 10^{-5}$ [13]. Such a discrepancy suggests that the effective damping constant of a virtual particle captured by a limited field of vision by the focused laser beam, whose velocity probability distribution obeys a peculiar non-Gaussian functions [Figs. 6, 7], is strongly modified from that of a real Brownian particle.

We have calculated the structure function by using long-term time series (> 10 seconds) of demodulated signals, and the Hurst exponent was found to decrease gradually and approach H = 0.5 (pure random walk) with increasing the observation time $\tau$, reflecting the corresponding approach toward asymptotic $f^{-2}$ behavior in the low frequency region shown in the power spectra. As for 107- and 207-nm particles in suspension, for example, the motion approaches pure random walks in the regime of $\tau > 3.3$ ms and $\tau > 10$ ms, respectively. An example result is shown in Fig. 9.

* Permanent address: *Department of Physics, Musashi Institute of Technology, Tamazutsumi, Setagaya, Tokyo 158-8557, Japan*